\documentclass[
	twocolumn,
]{aastex6} 

\usepackage{graphicx}
\usepackage{dcolumn}
\usepackage{bm}
\usepackage{color}
\usepackage{xspace}
\usepackage{ulem}
\newcommand{\fermi}{\emph{Fermi}\xspace}

\newcommand{\lat}{\emph{Fermi}-LAT\xspace}

\newcommand{\gr}{$\gamma$-ray\xspace}
\newcommand{\grs}{$\gamma$ rays\xspace}

\renewcommand{\deg}{\ensuremath{^{\circ}}\xspace}

\begin{document}
\title{Search for line-like and box-shaped spectral features from nearby galaxy clusters with 11.4 years of Fermi LAT data}

\author{Zhao-Qiang~Shen$^{1,\star}$, Zi-Qing~Xia$^{1}$, and Yi-Zhong~Fan$^{1,2,\star}$}
\affil{
	$^1$ {Key Laboratory of Dark Matter and Space Astronomy, Purple Mountain Observatory, Chinese Academy of Sciences, Nanjing 210008, China.}\\
	$^2$ {School of Astronomy and Space Science, University of Science and Technology of China, Hefei, Anhui 230026, China.}
}
\email{zqshen@pmo.ac.cn (ZQS), yzfan@pmo.ac.cn (YZF)}

\begin{abstract}
	Sharp spectral structures in the $\gamma$-ray band are an important dark matter (DM) signature.
	Previously, a tentative line feature at $\sim 43~{\rm GeV}$ is reported in 16 nearby galaxy clusters (GCls) with 7.1 years of \emph{Fermi}-LAT data, whose TS value is $\sim 16.7$.
	In this work, we search for line signals and box-shaped structures using the stacked data from those 16 GCls with 11.4-yr P8R3 data.
	There is still a hint at $\sim {42~\rm GeV}$, dominated by the radiation of Virgo and Ophiuchus clusters.
	Though the TS value was high up to 21.2 in October 2016, currently it has dropped to 13.1.
	Moreover, the TS value at $\sim {42~\rm GeV}$ decreases to 2.4 when the EDISP2 data are excluded from the analysis.
	Consequently, we do not find any statistically significant line-like signal and then set up the 95\% confidence level upper limits on the thermally averaged cross section of DM annihilating into double photons.
	The same line search has been carried out for an alternative GCl sample from the Two Micron All-Sky Survey but no any evidence has been found.
	We also search for box-shaped features in those 16 baseline GCls.
	No signal is found as well and the corresponding upper limits on the annihilation cross section are given.
\end{abstract}

\maketitle

\section{Introduction}\label{sec:introduction}
Observations require dark matter (DM) to explain various gravitational phenomena of different scales~\citep{Bertone2005,Bertone2018}.
According to the latest cosmic microwave background power spectrum measurement~\citep{Aghanim2018}, around 84.4\% of the matter density is contributed by DM.
Lots of DM candidates are proposed in literature~\citep{Feng2010}, and the weakly interacting massive particles (WIMPs) are the leading one since they can naturally explain today's relic density of DM.
If WIMPs annihilate or decay into standard model particles, the stable products may be observed by the \gr or cosmic ray detectors, thereby the particle nature can be inferred.
In many cases, the energy spectra of the products are smooth~\citep{Cirelli2011}, so it is often hard to distinguish them from astrophysical backgrounds (see~\citep{Charles2016} for a review).
However, if the products have a sharp feature, it will have a much better signal-to-noise ratio and therefore is more likely to be identified.

One of the sharp structures is the line-like signal, which can be produced by the annihilation or decay of WIMPs $\chi$ into a two-body final state $\gamma X$~\citep{Bergstrom1988}.
The \gr line is located at $E_\gamma=m_\chi (1-m_X^2/4m_\chi^2)$ for DM annihilation.
Such process is proposed in some extensions of standard particle physics models, e.g. the annihilation of the lightest neutralinos through $\chi\chi \to \gamma\gamma$ or $\chi\chi \to \gamma Z^0$~\citep{Rudaz1991}, or the decay of the gravitinos through $\chi \to \gamma \nu$~\citep{Ibarra2008}.

Great efforts have been paid to search for lines with \fermi Large Area Telescope (LAT)~\citep{FermiLAT2009}, but no significant signal has been found up to now~\citep{Abdo2010,Vertongen2011,Ackermann2012,Bringmann2012,Weniger2012,Tempel2012,Su2012a,Su2012,Huang2012,Tempel2012b,Ackermann2013a,Hektor2013,Albert2014,Ackermann2015,Anderson2016,Liang2016,Liang2016b,QuincyAdams2016,Liang2017,LiS2019}.
Interestingly, some tentative lines have been reported above $3\sigma$.
One is the $\sim 130~\rm GeV$ line first suggested in the Galactic halo using 3.6 years of \lat P7V6 data.
After considering the look-elsewhere effect, it still has a significance of $3.1\sigma$~\citep{Bringmann2012,Weniger2012}.
This line candidate was later confirmed by other groups~\citep{Tempel2012,Su2012a,Ackermann2013a} and also observed in unassociated sources~\citep{Su2012} and some Galaxy clusters (GCls)~\citep{Hektor2013}.
A weak excess at $\sim 110~\rm GeV$ was also reported in~\citet{Su2012} and \citet{Tempel2012b}.
However, this candidate reduced to $\sim 0.7\sigma$ in LAT P8 data and is consistent with the Earth Limb control region, which suggested a small systematic effect at this energy~\citep{Ackermann2015}.
Another tentative line was found at $\sim 43~\rm GeV$ in a GCls sample with a global significance of $3.0\sigma$ using 7.1-yr LAT P8R2 data~\citep{Liang2016}.
However, it does not appear in the Galactic center~\citep{Ackermann2015}, Milky Way satellites~\citep{Liang2016b} and subhalos~\citep{Liang2017,LiS2019}.
\gr line searches have also been performed with other telescopes such as DAMPE~\citep{Shen2019}, H.E.S.S.~\citep{Abdallah2018a,Abdalla2018b} and HAWC~\citep{Albert2020}, but only null results have been reported.

The box-shaped spectrum is also a prominent structure, which can be produced by the decay into double \grs of a pair of intermediate particle $\varphi$ from DM annihilation.
The intermediate particles can be standard model particles~\citep{Ibarra2012} or exotic particles such as axions~\citep{Ibarra2013}.
Box-shaped structures have been searched in the Galactic center~\citep{Ibarra2012,Ibarra2013}, dwarf spheroidal galaxies (dSphs)~\citep{LiS2018} and the Sun~\citep{Mazziotta2020} with \lat, but no signal is found either.

GCls are promising targets for DM indirect search, because they are the most massive gravitational bound systems in the Universe and contain a large number of substructures.
Unexpected continuum is search with \lat~\citep{Ackermann2010,Dugger2010,Huang2012b,Lisanti2018} and null results are obtained.
As previously mentioned, a tentative line was found in a GCl sample but not detected in other types of sources.
It may be possible considering the large uncertainty in the boost factor~\citep{Gao2012}.
In order to improve our understanding of this potential feature, we use 11.4 years of \lat data to revist the analysis.
We also investigate the temporal behavior of the line candidate.
Furthermore, we use another GCl sample from the Two Micron All-Sky Survey (2MASS)~\citep{Tully2015,Kourkchi2017,Lisanti2018} to check whether the line-like structure also exists.
The box-shaped structure is another interesting sharp structure, so we search for it in this work as well.

This paper is arranged as follows.
In Sect.~\ref{sec::data_analysis}, the baseline GCl sample, the \lat data and the likelihood method will be introduced.
We will then focus on the searches of line-like and box-shaped structures in Sect.~\ref{sec::line_search} and Sect.~\ref{sec::box_search} respectively.
Finally, a summary will be presented in Sect.~\ref{sec::summary}.

\section{Data analysis}\label{sec::data_analysis}
\subsection{galaxy clusters}
Our baseline sample of GCls is the same as that used in~\citet{Anderson2016} and \citet{Liang2016}.
This sample consists of 16 GCls with large $J$-factors, as listed in Tab.~\ref{tab:HIFLUGCS_GCls}.
15 sources of them are selected from the extended HIFLUGCS catalog~\citep{Reiprich2002,Chen2007}, whose masses can be easily measured with X-ray observations, and the other one is the Virgo cluster.
The positions ($\alpha$, $\delta$), redshifts $z$, virial radii $R_{200}$ and masses $M_{200}$ of these sources are directly taken from~\citet{Anderson2016}, where the $M_{200}$ and $R_{200}$ of HIFLUGCS GCls are derived from the $M_{500}$ and $R_{500}$ reported in~\citet{Reiprich2002} and \citet{Chen2007}, while the information of Virgo is taken from~\citet{Fouque2001}.
The region of interest (ROI) $\theta_{\rm ROI}$ is defined with the angular radius $\theta_{200}\equiv \arctan (R_{200}/d_A)$ for the sources except Virgo and M49, where $d_A$ is the angular diameter distance.
$2.6\deg$ and $1.7\deg$ are chosen as the size of ROIs for Virgo and M49 respectively to avoid overlapping~\citep{Liang2016}.

We assume that the smooth DM halo of a GCl follows the Navarro-Frenk-White (NFW) profile~\citep{Navarro1997}, i.e. $\rho_{\rm sm}(r) = \rho_0 / [(r/r_{\rm s})(1+r/r_{\rm s})^2]$.
With the help of the concentration relation in~\citet{SanchezConde2014}, we can derive the concentration parameter $c_{200}=R_{200}/r_{\rm s}$ with the $M_{200}$.
We then obtain the scale radius $r_{\rm s}$ using $R_{200}$, and the normalization $\rho_0$ using $M_{200}$.
Cosmological parameters~\citep{Aghanim2018} is adopted in the calculations.\footnote{If $H_0=74.03~\rm km\,s^{-1}\,Mpc^{-1}$~\citep{Riess2019} is used, the $J$-factors will increase by $\sim 75\%$.}
The $J$-factor of a smooth halo within a given ROI can be calculated with
\begin{equation}
	J_{\rm ROI}=\int_{\rm ROI} {\rm d} \Omega \int_0^\infty {\rm d}l\,\rho_{\rm sm}^2(r(l)),
\end{equation}
where $l$ is the line of sight distance from the Earth.

The substructures will enhance the annihilation of DM particles.
However, because of the uncertainties of the concentration relation and the mass function for the low-mass subhalos, the boost factor can vary from $\sim 3$ to $1000$~\citep{Gao2012,SanchezConde2014,Bartels2015,Ando2019}.
We adopt the relation given in~\citet{SanchezConde2014} since it gives an intermediate value.
For Virgo and M49 whose ROIs are smaller than their angular radii, the surface profile of subhalos in~\citet{Gao2012} is used to calculate the boost factors within the ROIs.
The average boost factor of our sample is $\bar{b}_{\rm sh,ROI}=\sum_i J_i\, b_{{\rm sh,ROI},i}/\sum_i J_i=24.7$.

\begin{table*}
	\centering
	\caption{\label{tab:HIFLUGCS_GCls}
	Parameters of the GCls in the baseline sample.
	The right ascension $\alpha$, declination $\delta$, redshift $z$, virial radius $R_{200}$ and mass $M_{200}$ of each GCl is taken from~\citet{Anderson2016}.
	$\theta_{\rm ROI}$ is the size of ROI from which we select the events.
	$J_{\rm ROI}$ and $b_{\rm sh,ROI}$ are the $J$-factors and subhalo boost factor inside the ROI respectively.
	}
	\begin{tabular}{lrrcccccc}
		\hline\hline
		GCl & $\alpha$ & $\delta$ &  $z$  &     $M_{200}$      & $R_{200}$ & $\theta_{\rm ROI}$ &  $\lg (J_{\rm ROI})$   &$b_{\rm sh,ROI}$ \\
		    & ($\deg$) & ($\deg$) &       & ($10^{14}M_\odot$) &($\rm Mpc$)&       ($\deg$)     & ($\rm GeV^2\,cm^{-5}\,sr$) &              \\
		\hline
		Virgo    &$ 187.70 $&$ 12.34 $&$0.004$&$ 5.60 $&$ 1.70 $&$ 2.60 $&$ 18.44 $&$ 18.6 $\\
		M49      &$ 187.40 $&$  8.00 $&$0.003$&$ 0.72 $&$ 0.88 $&$ 1.70 $&$ 17.88 $&$ 14.5 $\\
		Ophiuchus&$ 258.10 $&$-23.38 $&$0.028$&$42.44 $&$ 3.43 $&$ 1.63 $&$ 17.73 $&$ 36.2 $\\
		Fornax   &$  54.63 $&$-35.45 $&$0.005$&$ 1.39 $&$ 1.10 $&$ 2.84 $&$ 17.72 $&$ 30.1 $\\
		A3526    &$ 192.20 $&$-41.31 $&$0.011$&$ 3.72 $&$ 1.52 $&$ 1.80 $&$ 17.44 $&$ 33.0 $\\
		NGC~4636 &$ 190.70 $&$  2.69 $&$0.003$&$ 0.19 $&$ 0.56 $&$ 2.43 $&$ 17.36 $&$ 23.4 $\\
		S636     &$ 157.50 $&$-35.32 $&$0.009$&$ 1.69 $&$ 1.17 $&$ 1.69 $&$ 17.29 $&$ 30.8 $\\
		Coma     &$ 195.00 $&$ 27.98 $&$0.023$&$10.92 $&$ 2.18 $&$ 1.25 $&$ 17.28 $&$ 35.3 $\\
		A3627    &$ 243.90 $&$-60.91 $&$0.016$&$ 5.38 $&$ 1.72 $&$ 1.41 $&$ 17.28 $&$ 34.0 $\\
		Perseus  &$  49.65 $&$ 41.52 $&$0.018$&$ 6.66 $&$ 1.85 $&$ 1.35 $&$ 17.27 $&$ 34.4 $\\
		AWM7     &$  43.63 $&$ 41.59 $&$0.017$&$ 5.38 $&$ 1.72 $&$ 1.33 $&$ 17.23 $&$ 34.0 $\\
		A1367    &$ 176.10 $&$ 19.84 $&$0.022$&$ 8.13 $&$ 1.98 $&$ 1.19 $&$ 17.20 $&$ 34.9 $\\
		A1060    &$ 159.20 $&$-27.53 $&$0.013$&$ 2.72 $&$ 1.37 $&$ 1.38 $&$ 17.17 $&$ 32.2 $\\
		3C~129   &$  72.29 $&$ 45.01 $&$0.021$&$ 5.90 $&$ 1.78 $&$ 1.12 $&$ 17.10 $&$ 34.2 $\\
		A2877    &$  17.45 $&$-45.90 $&$0.025$&$ 7.54 $&$ 1.93 $&$ 1.02 $&$ 17.06 $&$ 34.7 $\\
		NGC~5813 &$ 225.30 $&$  1.70 $&$0.007$&$ 0.46 $&$ 0.76 $&$ 1.40 $&$ 16.99 $&$ 26.4 $\\
		\hline\hline
	\end{tabular}
\end{table*}

\subsection{$\gamma$-ray data}
\lat P8R3 data is based on the most recent event-level analysis, which alleviates the background cosmic rays leaked from the ribbons of the anti-coincidence detector~\citep{Bruel2018}.
To further reduce the cosmic-ray contamination in the data, we only use the events satisfying the {\tt ULTRACLEAN} selection~\citep{Ackermann2012_CKWEC}.
According to the reconstruction quality of energy, P8 data are further classified into four disjoint subsets.
They are denoted by EDISP0--3, each accounting for a quarter of data, according to the energy reconstruction quality estimator~\citep{Atwood2013_C82D0}.
Since the energy resolution of EDISP0 data is significantly worse than that of the remaining events, we remove them to improve the sensitivity.
Unless otherwise noted, the analysis is based on the combined data set of EDISP1+EDISP2+EDISP3.

The data collected between 27 October 2008 and 19 March 2020 (\fermi Mission Elapsing Time between 246823875 and 606273279) are selected.\footnote{\url{https://heasarc.gsfc.nasa.gov/FTP/fermi/data/lat/weekly/photon/}.}
To reduce the emission from the Earth Limb, we only use the events with zenith-angle less than $90\deg$.
In addition, we apply the quality-filter cut $\rm (DATA\_QUAL==1)\&\&(LAT\_CONFIG==1)$ which removes the time intervals and the corresponding events when \lat is not the science mode or when strong solar flares or particle events happens.

\begin{figure}
    \centering
	\includegraphics[width=0.48\textwidth]{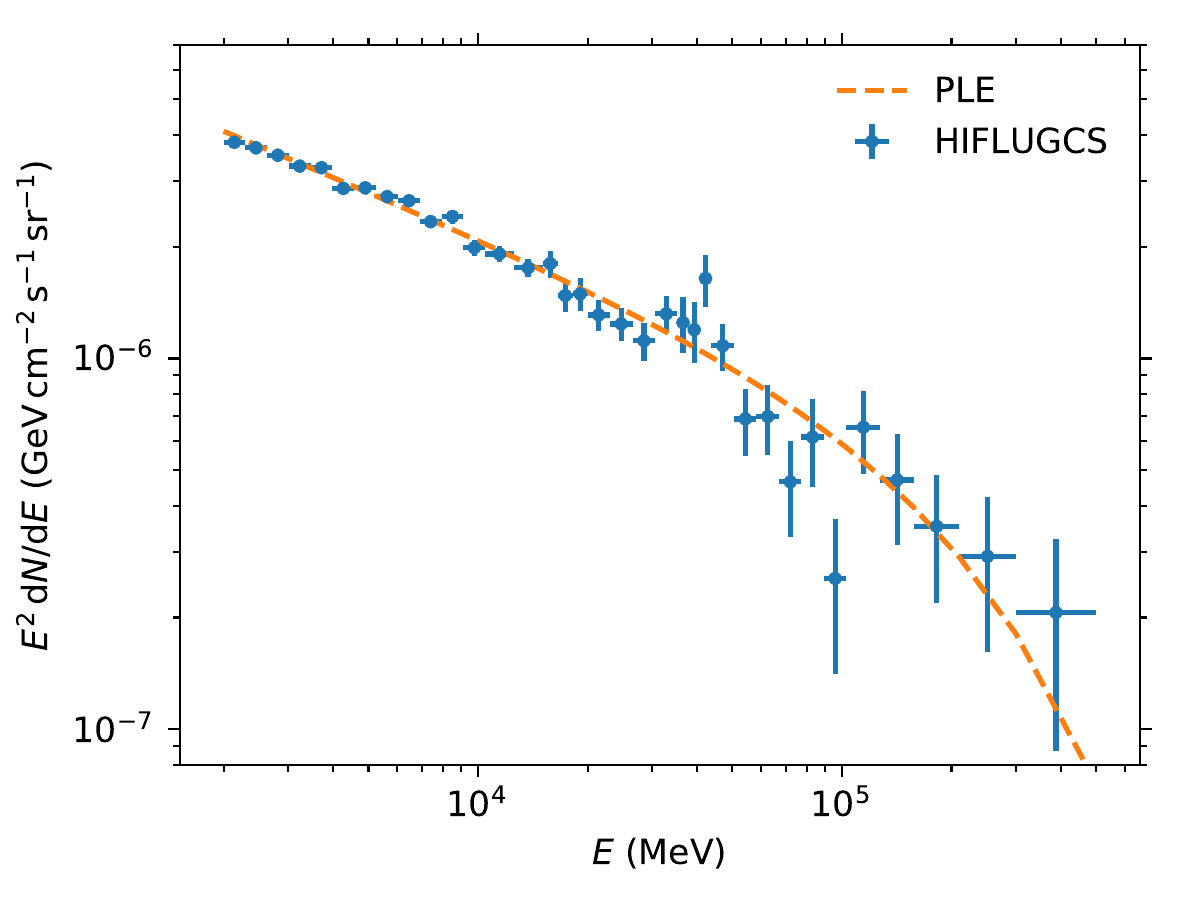}
    \caption{\label{fig:flux}
		The stacked \gr spectral energy distribution (SED) of the 16 GCls in the baseline sample.
		The dashed line is the best-fit powerlaw with exponential cutoff (PLE) model using the unbinned likelihood defined in Sect.~\ref{sec::data_analysis:unbinned_like}.
    }
\end{figure}

We use events within the ROIs to perform the analysis.
The stacked spectrum energy distribution (SED) shown in Fig.~\ref{fig:flux} is derived with
\begin{equation}
	({\rm d}N/{\rm d}E)_j = \frac{ \sum_i n_{ij}}{\Delta E_j\,\sum_i \Omega_i\,\epsilon_{ij}},
\end{equation}
where for the $i$-th GCl, $n_{ij}$ is the number of photons in the energy bin centering at $E_j$, while the $\Omega_i$ and $\epsilon_{ij}$ are the solid angle and the average exposure respectively.
We utilize the {\tt Fermitools v1.0.10}\footnote{\url{https://fermi.gsfc.nasa.gov/ssc/data/analysis/software/}.} to select the events and make the live time cube.

An excess at $\sim 43~\rm GeV$ is still distinct comparing to the best-fit power law with exponential cutoff (PLE) model.
To derive the significance and constraints of this structure, we adopt the unbinned likelihood method in the following sections, which alleviates the uncertainty caused by the energy binning.

\subsection{unbinned likelihood}\label{sec::data_analysis:unbinned_like}
The unbinned likelihood method is used to derive the significance of this suspected line and the constraints of the annihilation cross section.
The events of all the target sources within their ROIs are put together in the fitting to improve the sensitivity.
The background in the stacking data is contributed by various point sources, the Galactic diffuse emission and the isotropic background.
To reduce the influence of an inaccurate background spectral shape on the signal, we adopt the sliding windows technique:
fittings are only performed using the data within a given small energy bin $[E_{\min},E_{\max}]$ containing the signal.
In such an energy window, it is safer to choose a powerlaw model as an approximation of the background spectrum.
We set $E_{\min}=E_- - 0.5E_{\rm c}$ and $E_{\max}=E_+ + 0.5E_{\rm c}$, where $E_\pm$ is the bounds of the sharp structure ($E_+=E_-$ for a line), and $E_{\rm c}=(E_++E_-)/2$.
The energy difference between two adjacent windows is $0.5\sigma(E_{\rm c})$, where $\sigma(E)$ is the half width of 68\% energy dispersion containment at the energy $E$.

The unbinned likelihood function $L$ in the $k$-th energy window is defined as~\citep{Mattox1996,Ackermann2013a,Liang2016}
\begin{equation}
	\ln L_k(\Theta) = - \int_{E_{{\rm min},k}}^{E_{{\rm max},k}} \lambda(E;\Theta)\,{\rm d}E + \sum_{j=1}^{N_{{\rm ph},k}} \ln [\lambda(E_j)].
\end{equation}
$N_{{\rm ph},k}$ is number of photons in the window.
$\lambda(E;\Theta)$ is the expected counts of the target sources per energy range with the parameter $\Theta$.
Hypothesis tests are performed to evaluate the pre-trial significance of lines.
For the null hypothesis, photons are only contributed by the powerlaw background.
Therefore we have
\begin{equation}
	\lambda_{\rm null}(E;N_{\rm b},\Gamma) = N_{\rm b}\,E^{-\Gamma} \sum\nolimits_i \epsilon_i(E)\,\Omega_i,
\end{equation}
where $i$ is the index of a GCl.
For the alternative hypothesis, the \gr model is made of a powerlaw background and a sharp structure and is given by
\begin{eqnarray}
	\label{eqn:lambda_alt}
	\lambda_{\rm alt}
		&=& \sum\nolimits_i \Omega_i \int\int {\rm d}E'{\rm d}\theta\, \phi(E';N_{\rm s})\,D_i(E;E',\theta)\,\epsilon_i(E',\theta) \nonumber\\
		&+& \lambda_{\rm null}(E;N_{\rm b},\Gamma).
\end{eqnarray}
$D(E;E',\theta)$ is the energy dispersion function,\footnote{\url{https://fermi.gsfc.nasa.gov/ssc/data/analysis/documentation/Cicerone/Cicerone_LAT_IRFs/IRF_E_dispersion.html}}
which is related to the incident angle $\theta$ and energy $E'$.
$\phi(E)$ can be the spectrum of the line-like or box-like structure.

For a line signal caused by the direct annihilation of DM particles into photon pairs, we have $\phi(E)=N_{\rm s}\,\delta(E-E_{\rm line})$, where $E_{\rm line}$ is the line's energy and also the DM mass $m_\chi$.
$\delta(x)$ is the delta function.
$N_{\rm s}$ is the normalization factor.
In this case, the first part of Eq.(\ref{eqn:lambda_alt}) can be reduced to $N_{\rm s}\sum\nolimits_i \Omega_i \int {\rm d}\theta\, D_i(E;E_{\rm line}, \theta)\,\epsilon_i(E_{\rm line}, \theta)$.

For a box-shaped signal caused by the process $\chi\chi \to \varphi\varphi$ and $\varphi \to \gamma\gamma$, we define the spectrum as $\phi(E)=N_{\rm s}\,H(E - E_-)\,H(E_+ - E)/\Delta E$~\citep{Ibarra2012}.
$H(x)$ is the Heaviside step function.
$E_\pm = (m_\chi/2)(1\pm \sqrt{1-m_\varphi^2/m_\chi^2})$ are the lower and upper bounds of the spectrum.
$\Delta E=E_+-E_-$ is its width, which is related to the degeneracy parameter $\Delta m/m_\chi \equiv (m_\chi - m_\varphi)/m_\chi$~\citep{Ibarra2012}.

We optimize the likelihood function with the {\tt MINUIT} algorithm~\citep{MINUIT1975}.
The likelihood ratio test is performed to find the goodness of the alternative model.
We use the likelihood ratio test statistic ${\rm TS}=-2 \ln \Lambda$, where $\Lambda \equiv \hat{L}_{\rm null}/\hat{L}_{\rm sig}$ is the likelihood ratio of the best-fit null and alternative models.
According to the Chernoff's theorem~\citep{Chernoff1954}, the TS value distributes according to $\frac{1}{2} (\delta+\chi^2_{\rm dof=1})$.
If no significant signal is detected, the 95\% confidence level upper limit can be derived by increasing $N_{\rm s}$ until the $\ln L_{\rm sig}$ decreases by 1.35 (2.71/2) with respect to the maximum value.

\section{Search for line-like structures}\label{sec::line_search}

\begin{figure*}
    \centering
	\includegraphics[width=0.48\textwidth]{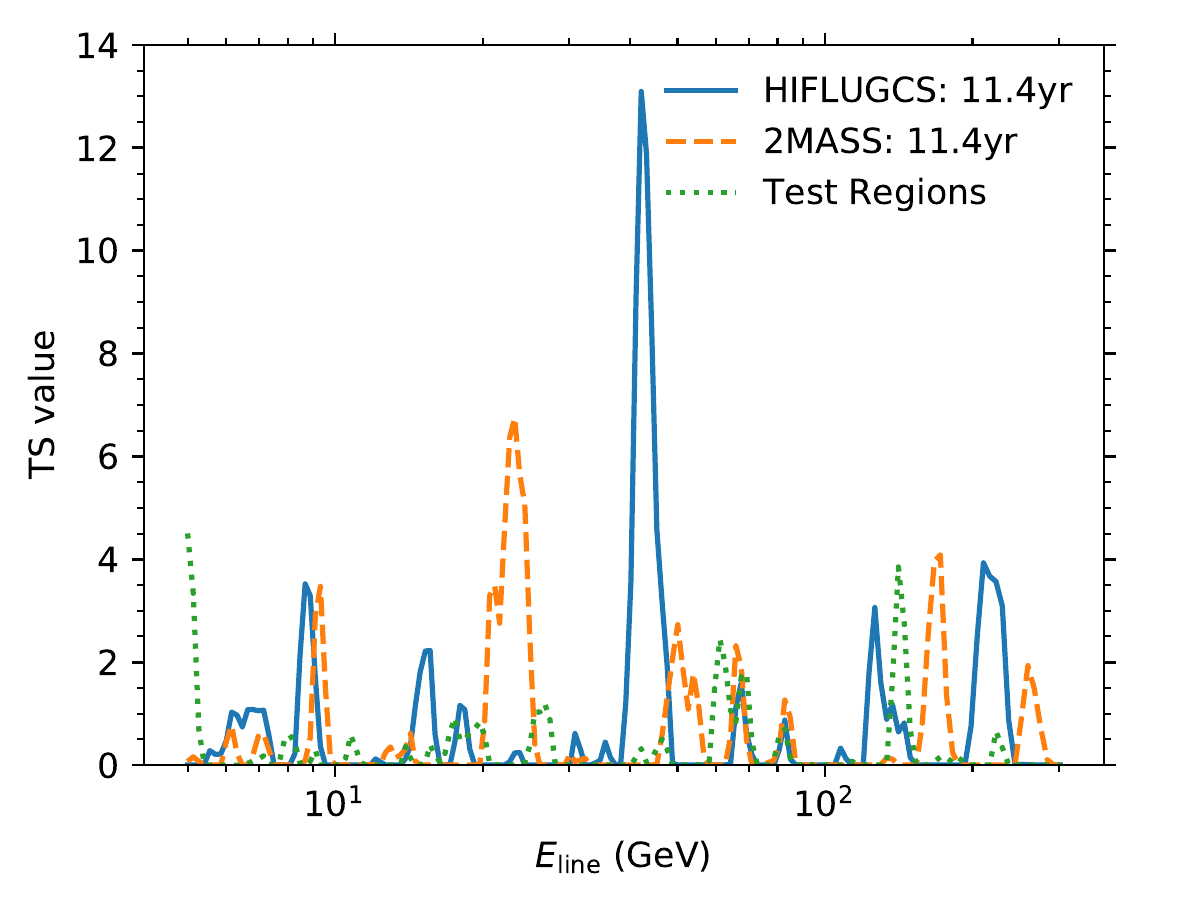}
	\includegraphics[width=0.48\textwidth]{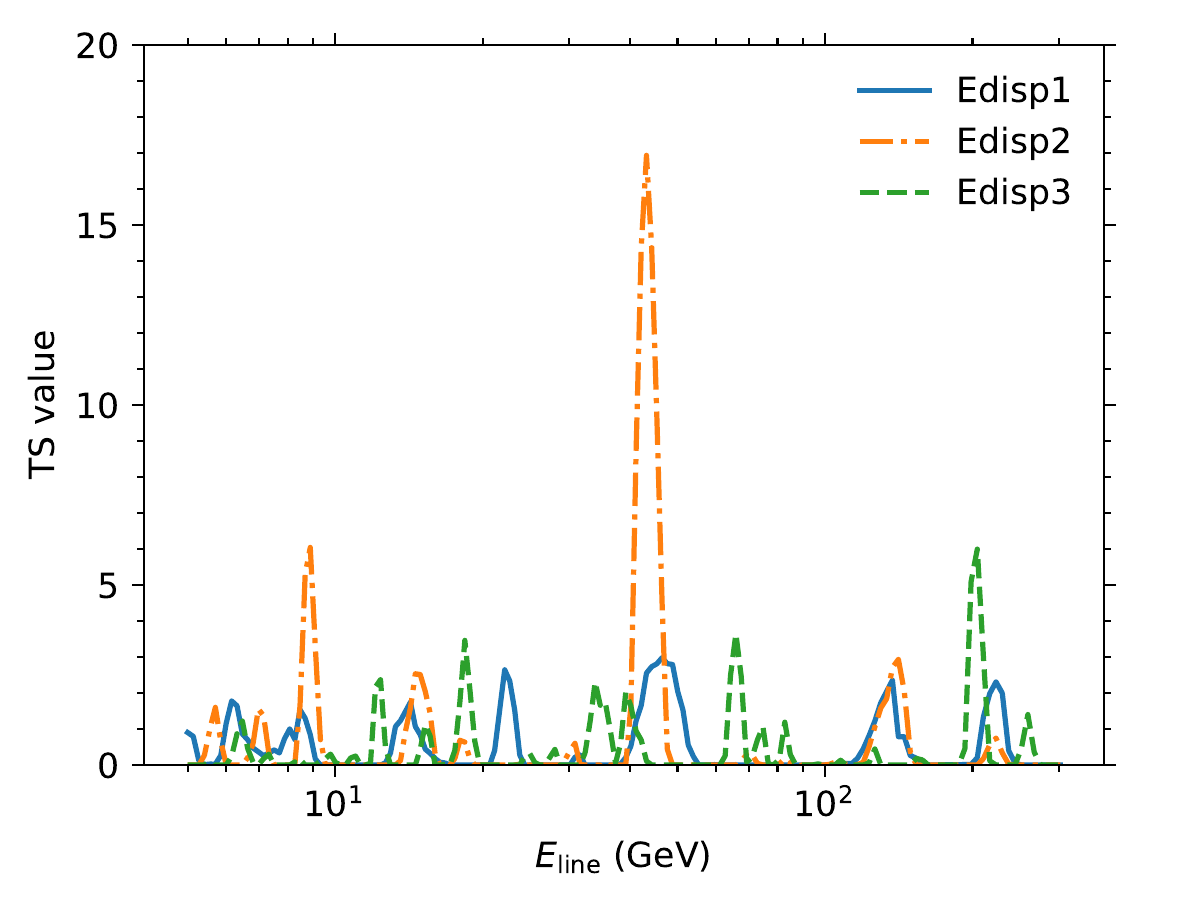}
    \caption{\label{fig:ts}
		The TS values of possible line signatures at different energies.
		The left and right panels show the TS values achieved with different GCl samples and different event types of photons from the HIFLUGCS sample, respectively. 
    }
\end{figure*}

\begin{figure}
    \centering
	\includegraphics[width=0.48\textwidth]{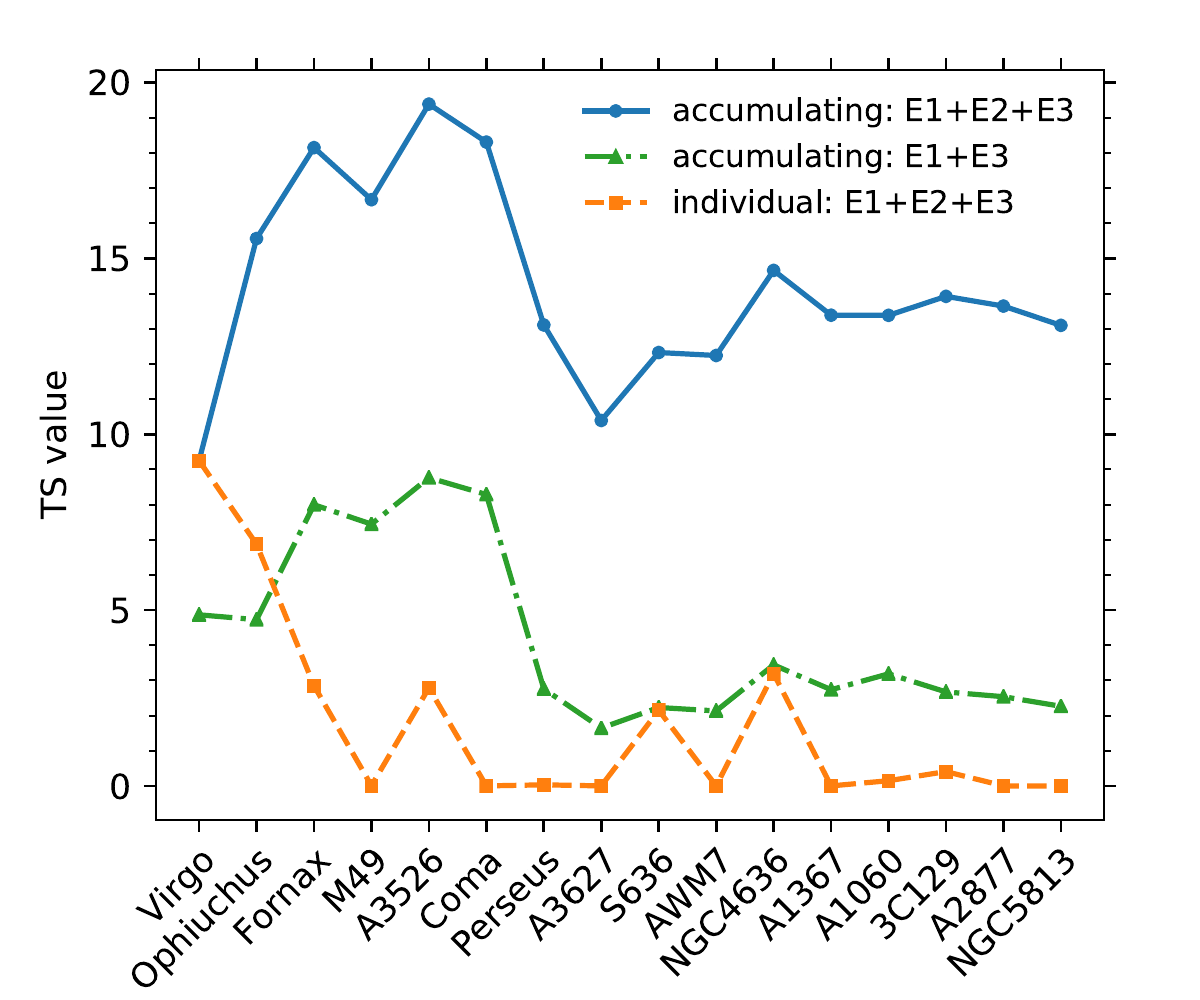}
    \caption{\label{fig:ts_nsrc}
		The TS values of 42.2 GeV line derived with individual HIFLUGCS GCl (dashed orange) and with the gradually increasing GCl sample (blue solid).
		The GCls labelled in the $x$-axis are sorted according to the boosted $J$-factors within the ROIs.
		The TS values excluding the EDISP2 data are also shown in a green dot-dashed line.
    }
\end{figure}

\begin{figure}
	\centering
	\includegraphics[width=0.48\textwidth]{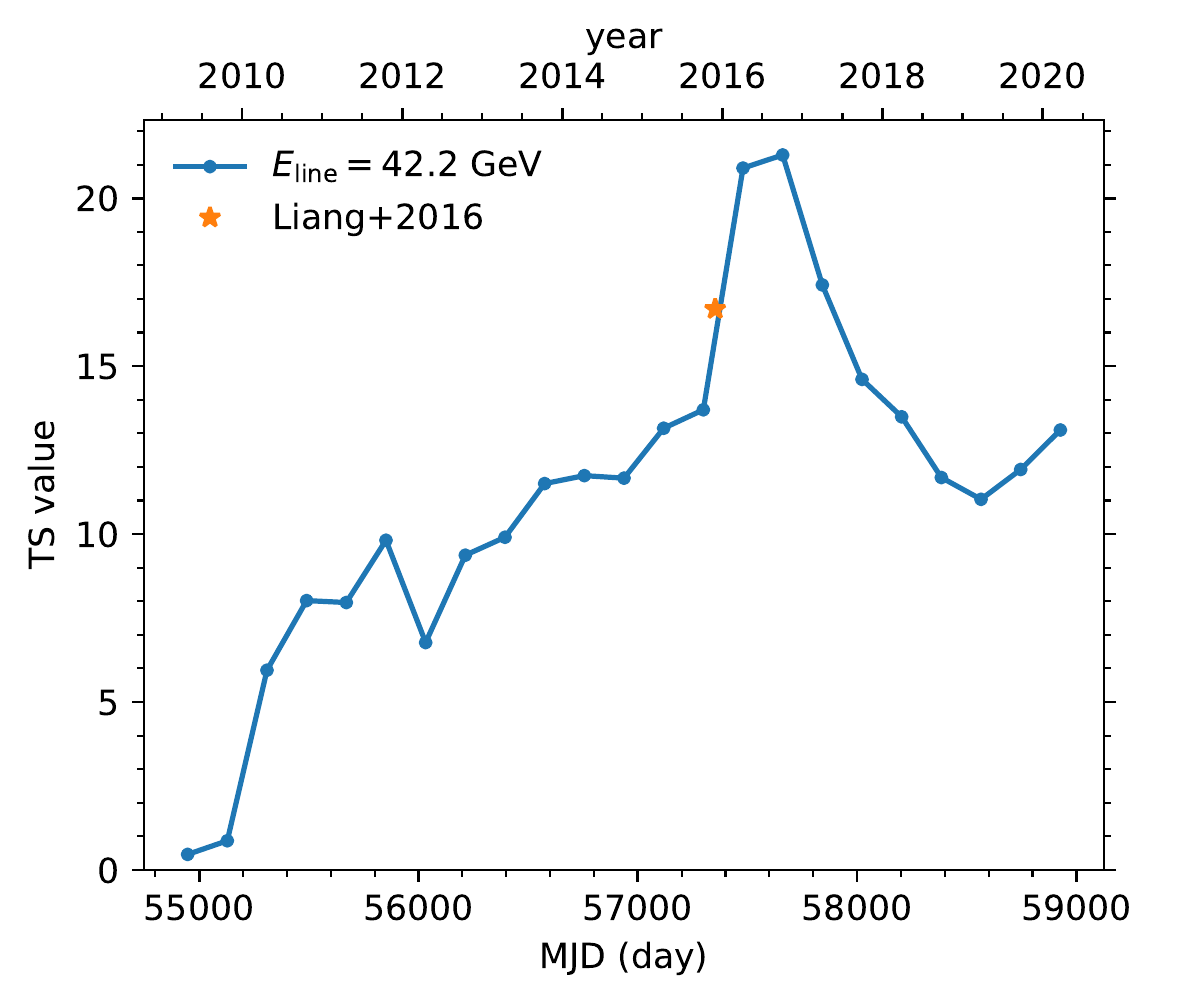}
    \caption{\label{fig:ts_lc}
		The change of TS values of the 42.2~GeV line with time using accumulating data.
		Each TS value is derived using the data from 27 October 2008 to the date given by the $x$-axis value.
		The orange star is the TS value presented in~\citet{Liang2016}.
    }
\end{figure}

The $\sim 43~\rm GeV$ line candidate was first reported with a global significance of $3.0\sigma$ using 7.1-yr \lat P8R2 data~\citep{Liang2016}.
Since $\sim 4$ more years of observations have been made by \lat and updated P8R3 data are released, we revisit the line search at $5-300~\rm GeV$.
In each window, we set the line energy fixed and fit the normalization of the line and spectral parameters of the background.
The TS values of lines at different energies are shown in the left panel of Fig.~\ref{fig:ts}.
Clearly, no significant line signal is detected.
The largest TS value is 13.1 at the energy of $\sim 42.2~\rm GeV$.
Since the pre-trial/local significance of a line is $s_{\rm local}=\sqrt{\rm TS}$~\citep{Ackermann2013a}, the TS value corresponds to a local significance of $3.6\sigma$.
Considering 40.9 independent trials are made when performing the sliding windows technique~\citep{Liang2016}, the global significance is $2.5\sigma$.

To look into this issue, we split the data according to the event types and repeat the unbinned analyses respectively.
As shown in the right panel of Fig.~\ref{fig:ts}, the $\sim 42~\rm GeV$ spike only displays in the EDISP2 data with a TS value of 16.9, but absent in both the EDISP1 and EDISP3 data, which might indicate an instrumental origin.
But anyway both the results with and without EDISP2 data will be presented in the paper, and the constraints are derived with EDISP2 data included in order to be conservative.

In addition, as shown in Fig.~\ref{fig:ts_nsrc}, we calculate the individual TS value for each cluster and the accumulating TS value for the gradually increasing GCl samples.
If EDISP2 data are included in the data, the GCls with large $J$-factors appear to have large TS values.
Particularly the TS values of the line are 9.3 and 6.9 for Virgo and Ophiuchus, respectively.
If we stack the data from the top 5 GCls, the TS value is 19.4, which means a global significance of $2.7\sigma$ with a trial factor of 654.4.
The tentative line at $\sim 42$ GeV happens to appear in all the top 7 GCls with the largest unboosted $J$-factors except M49.
The stacked TS value of these 7 GCls is 24.6, corresponding to a $3.5\sigma$ global significance.
When the EDISP2 data are dropped, the TS value decreases to 8.7 for the top 5 GCls, but is still around 5.0 in the Virgo cluster.

Analyses are also done in ``test regions" around the GCls.
Firstly, since the Virgo cluster has the largest TS value, we define a test region around it with a ring-shaped region whose the inner and outer angular radii are 6.28\deg~\citep{Anderson2016} and 7.5\deg so as to remove the events associated with Virgo and keep the similar photon counts as the source region in the energy window.
A zero TS value is achieved in the test region when we carry out the same unbinned likelihood analysis.
To extend the analysis to the stacked GCls, we define the test ROIs whose centers are shifted twice the angular radii from the original positions of GCls to avoid overlapping.
Lines are also searched within these regions and the TS values are drawn in Fig.~\ref{fig:ts}.
Similarly, no lines with TS values larger than 5 are found in the test regions.

\begin{figure*}
	\centering
	\includegraphics[width=0.48\textwidth]{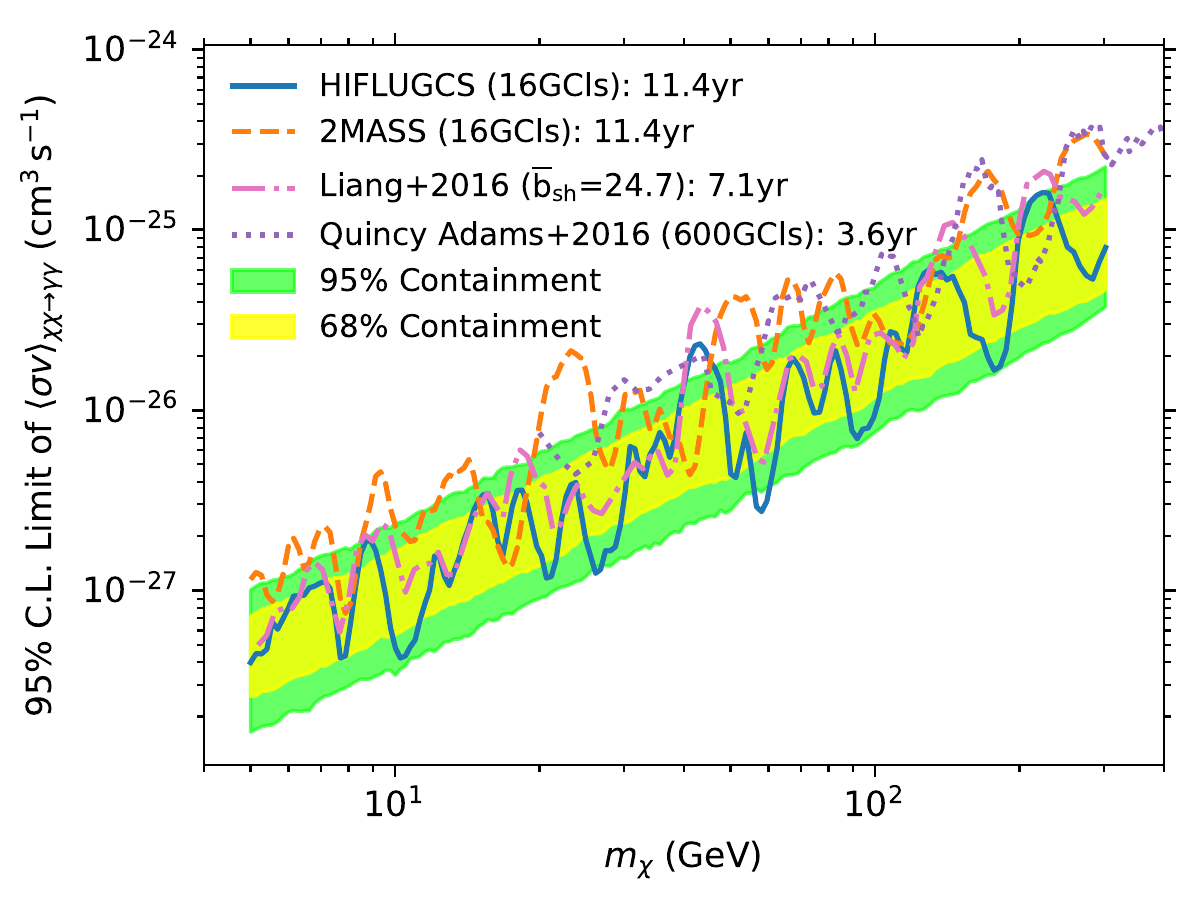}
	\includegraphics[width=0.48\textwidth]{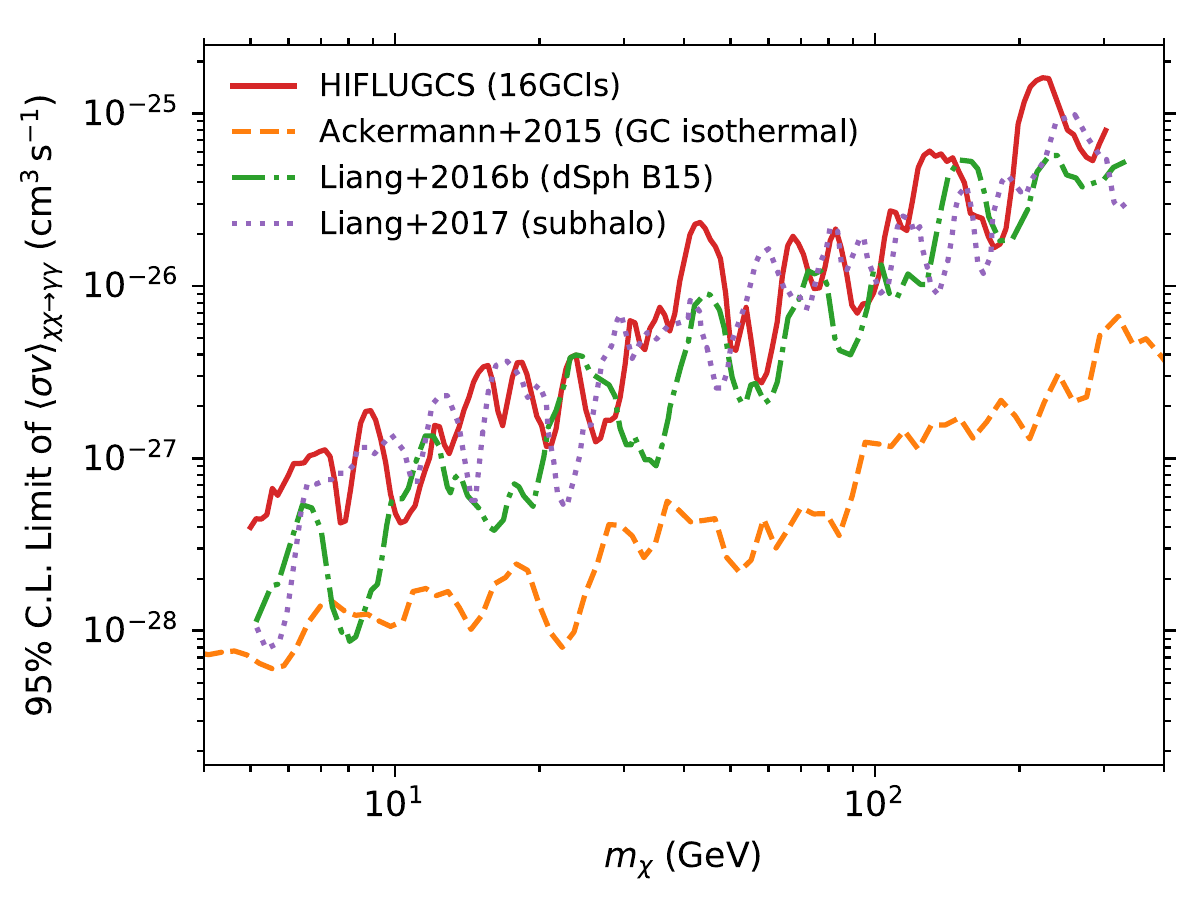}
    \caption{\label{fig:sv}
		The 95\% confidence level upper limit on the thermally averaged DM annihilation cross section into double \grs.
		The left panel shows the constraints from the Galaxy clusters.
		The blue solid line and orange dashed line represent the results derived with the baseline GCl sample and 2MASS GCl sample respectively.
		The pink dot-dashed and purple dotted lines are the GCl constraints from~\citet{Liang2016} and~\citet{QuincyAdams2016} respectively.
		Yellow and green band show the 68\% and 95\% containment of limits.
		The right panel shows the constraints from the Galaxy center (isothermal profile)~\citep{Ackermann2015}, dwarf galaxies (B15 sample)~\citep{Liang2016b} and the subhalo population~\citep{Liang2017}.
	}
\end{figure*}

The signal-to-background ratio is expected to grow as the square root of the observing time for a signal limited by the background emission.
However the latest TS value is smaller than the previous value of 16.7 reported in~\citet{Liang2016}.
To further investigate the change of TS value with the observation time, we split the observation equally into 23 parts with each one containing approximately half a year of data.
We gradually add each part to the fitting and perform the line search.
In this way, the TS values of the $42.2~\rm GeV$ line using accumulating data are achieved and are shown in Fig.~\ref{fig:ts_lc}.
The result of~\citet{Liang2016} is also marked in the figure with an orange star.
The TS value of the line has a growing trend before October 2016 when more data are included.
A TS peak of 21.2 is reached in October 2016 (7.9-yr data) and after then the significance begins to decrease.
The drop of TS value is caused by both the slight increase of the background emission and the decrease of the line signal.
The expected photon number of the line is $36.5\pm9.5$ in 7.9-yr data but drops to $34.2\pm9.8$ in 11.4-yr data according to the best-fit models.
If the count of the line signal increases proportionally to the observing time, this result may imply a $\sim 2\sigma$ tension.

If the annihilation of DM really happens and is strongly boosted in the GCls, the tentative line may also be found in an independent GCl sample.
The GCl catalog constructed from the 2MASS Redshift Survey~\citep{Tully2015,Kourkchi2017,Lisanti2018} is adopted to create a different GCl sample.
We first sort these GCls according to their $J$-factors, delete the Andromeda following~\citet{Lisanti2018}, and then remove those overlapping with the regions (defined with $\theta_{200}$) of other sources.
The GCls within $|b|<20\deg$ are also excluded to reduce the events from the Galactic plane~\citep{Lisanti2018}.
We select the top 16 GCls whose ROIs do not overlap with the baseline sample or the significant sources ($>1000\sigma$) in \lat 4FGL-DR2~\citep{4FGL2020}, where the ROIs is defined with $\theta_{200}/2$ of the GCls.\footnote{The ROI size $\theta_{200}/2$ reduces the probability of overlapping but has a small impact on the $J$-factor.}
The boost factors within the ROIs are calculated with~\citep{SanchezConde2014,Gao2012}.
The same unbinned likelihood method with the sliding windows technique is applied to the stacked photon events, and the resulting TS value is shown in Fig.~\ref{fig:ts}.
No lines including the 42.2 GeV line candidate are detected in this GCl sample.

The line flux induced by the annihilation of DM into photons is
\begin{equation}
	\phi(E) = \frac{\left< \sigma v \right>_{\chi\chi \to \gamma\gamma}}{8\pi m_\chi^2} 2\delta(E-E_{\rm line}) \times J_{\rm tot},
\end{equation}
where $\left < \sigma v \right>_{\chi\chi \to \gamma\gamma}$ is the thermally averaged cross section for $\chi\chi \to \gamma\gamma$ and $E_{\rm line}= m_\chi$.
$J_{\rm tot}$ is the sum of boosted $J$-factors of the GCls, which can be calculated with
\begin{equation}
	J_{\rm tot} = \sum_{i=1}^{16} J_{{\rm ROI},i}\times (1+b_{{\rm sh,ROI},i}).
\end{equation}
Since no signal is found, the 95\% confidence level upper limit on $\left < \sigma v \right>$ can be evaluated based on the upper limit of the normalization $N_{\rm s}$.
In Fig.~\ref{fig:sv}, we present our constraints using baseline GCl sample (blue solid line) and 2MASS sample (orange dashed line).
Also shown are the expected 68\% and 95\% containment of upper limits obtained with 1000 no-DM Monte Carlo simulations.
For the baseline sample, the fluctuation in the constraint is mostly within the 95\% containment region except that around 42~GeV.
It is reasonable since only the $\sim 42$~GeV candidate has a global significance larger than $2\sigma$.
On the other hand, the constraint from the 2MASS GCl sample restricts the parameter space of the line candidate since no tentative signal is found in that data set.
Yet, we are aware that the $J$-factors of 2MASS GCls have an uncertainty of $\sim 0.35~\rm dex$~\citep{Lisanti2018}.
We also plot the results from GCls~\citep{Liang2016,QuincyAdams2016}, the Galaxy center~\citep{Ackermann2015}, dwarf galaxies~\citep{Liang2016b}, and Milky way subhalos~\citep{Liang2017}.
Our result from the baseline sample is stronger than the previous GCls results, but is much weaker than the constraints from the Galactic center, dwarf galaxies and the subhalo population.

\section{Search for box-shaped structures}\label{sec::box_search}

\begin{figure}
	\centering
	\includegraphics[width=0.48\textwidth]{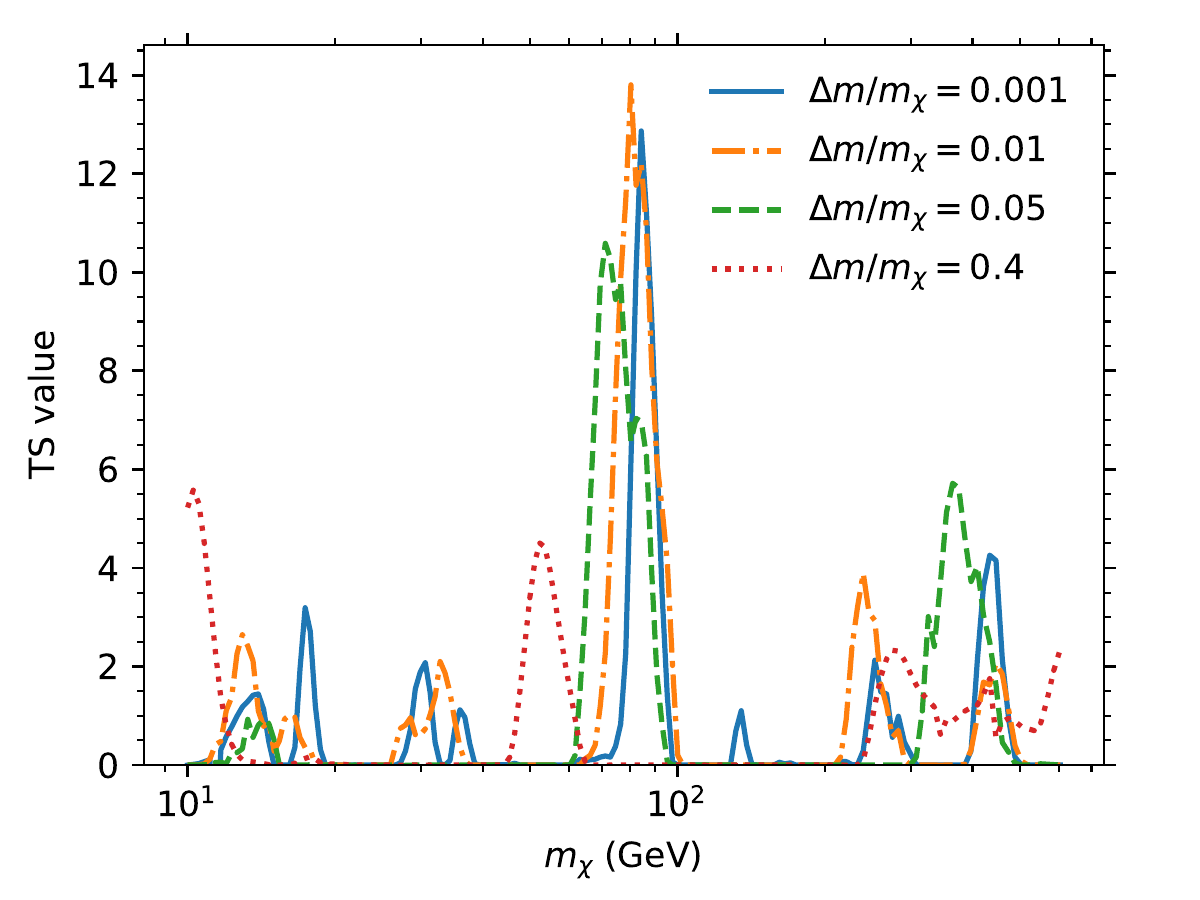}
    \caption{\label{fig:ts_box}
		The TS values of box-like signals at different energies for several $\Delta m/m_\chi \equiv (m_\chi - m_\varphi)/m_\chi$.
	}
\end{figure}

\begin{figure*}
	\centering
	\includegraphics[width=0.48\textwidth]{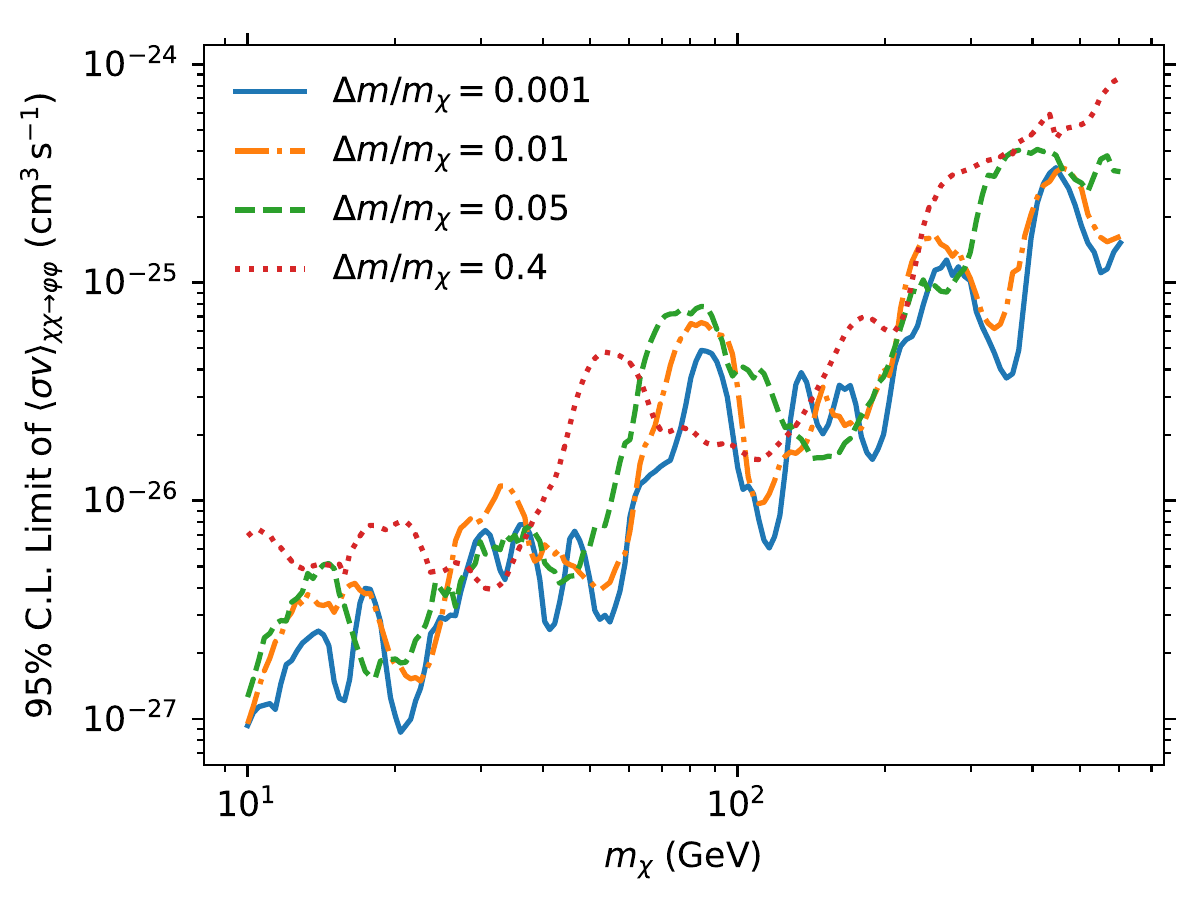}
	\includegraphics[width=0.48\textwidth]{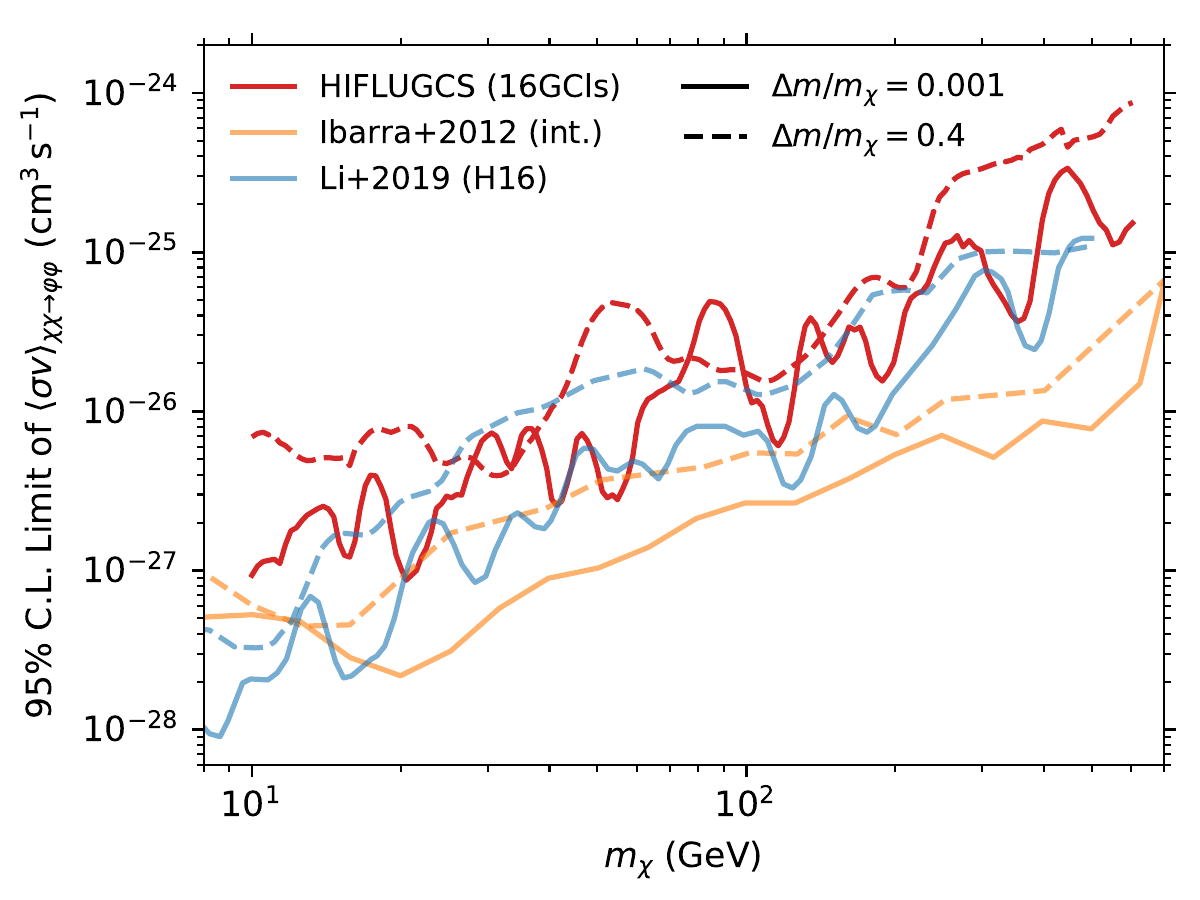}
    \caption{\label{fig:sv_box}
		The 95\% confidence level upper limit on the cross section of DM cascade annihilation using the baseline GCl sample when the intermediate particle $\varphi$ decays into two photons with the branching ratio of 1.
		The left panel shows the constraints for several degeneracy parameter $\Delta m/m_\chi$ values.
		The right panel compares our results with the previous constraints based on the data of Galactic center (intermediate approach)~\citep{Ibarra2012} and dSphs (H16 sample)~\citep{LiS2018} for $\Delta m/m_\chi=0.001$ and $0.4$.
	}
\end{figure*}

The box-shaped structures have been explored in the data of the Galactic center~\citep{Ibarra2012,Ibarra2013} and dSphs~\citep{LiS2018}.
In this section, we also search the GCls for the possible box-like excess.
We are mainly interested in the box-shaped spectrum which has a similar width as the energy dispersion, where the width is $\Delta E = E_+ - E_- = \sqrt{m_\chi ^2-m_\varphi^2}$.
Considering that the energy resolution $\Delta E/E$ is $\gtrsim 6\%$ for \lat P8R3 data, we take into account the box-shaped structures with the degeneracy parameter $\Delta m/m_\chi$ of 0.001, 0.01 and 0.05, which are approximately 1, 2 and 5 times as wide as the best energy resolution respectively.
Different from the previous works in~\citet{Ibarra2012,Ibarra2013} and \citet{LiS2018}, we convolve the narrow boxes with the energy dispersion functions provided by {\tt Fermitools} instead of Gaussian functions.
In addition, the unbinned likelihood method with the sliding windows technique is applied rather than performing a broad-band fit, which shares the same method as the line search above.
For a broad box-shaped feature, the constraint mainly comes from its high-energy edge and therefore can be converted between the boxes with different widths, so we only analyze the broad box with $\Delta m/m_\chi=0.4$ as a representative.
Different from the searching of narrow box-like spectra, we perform a broad-band fit with the events above 2.5~GeV, where the background emission is assumed to be a PLE model.

In Fig.~\ref{fig:ts_box}, we show the TS values of the box-shaped excess with respect to the DM mass.
Similar to the line search results before, we do not find any significant box-shaped structure.
A tentative excess is found at $m_\chi=84.3~\rm GeV$ (80.3~GeV, 71.2~GeV) with the maximum TS value being 12.9 (13.8, 10.6) for the boxes with $\Delta m/m_\chi=0.001$ (0.01, 0.05).
They are all caused by the sharp structure at $\sim 40~\rm GeV$ in the stacked SED.
The box with $\Delta m/m_\chi=0.01$ has a slightly larger TS value, which shows that the sharp component is wider than the energy resolution.
Moreover, no broad box-like candidate ($\Delta m/m_\chi=0.4$) is found with TS value larger than 9.
If we just analyze the EDISP1 and EDISP3 data, the TS values of the above signal decrease to 2.6, 4.2, and 3.1 for $\Delta m/m_\chi=0.001$, 0.01, and 0.05, respectively.

Because no significant box-shaped structures are found, the constraints are set.
We focus on the case when the branching ratio of $\varphi \to \gamma\gamma$ is 1.
The expected flux is given by~\citep{Ibarra2012}
\begin{equation}
	\phi(E) = \frac{\left< \sigma v \right>_{\chi\chi \to \varphi\varphi}}{8\pi m_\chi^2} \frac{4}{\Delta E} H(E - E_-)\,H(E_+ - E)\times J_{\rm tot},
\end{equation}
where $\left< \sigma v \right>_{\chi\chi \to \varphi\varphi}$ is the thermally averaged annihilation cross section into intermediate particle $\varphi$ pairs.
The left panel of Fig.~\ref{fig:sv_box} shows the constraints for several degeneracy parameters.
The upper limit of the narrowest box has a similar shape to that of the line except that the former is slightly smoother than the latter.
The boxes with larger $\Delta m/m_\chi$ have broader shape, hence have weaker and smoother constraints.
Comparing with the constraints from the Galactic center (orange lines)~\citep{Ibarra2012} and dSphs (blue lines)~\citep{LiS2018} for $\Delta m/m_\chi=0.001$ and $0.4$ in the right panel of Fig.~\ref{fig:sv_box}, our results are weaker if the average boost factor of $\sim 25$ is adopted.
However, better constraints could be achieved if the boost factors of GCls are larger than that adopted in this work.

\section{Summary}\label{sec::summary}
Sharp structures in the \gr spectra can be related to the annihilation or decay of DM particles.
In this work, we use 11.4-yr \lat P8R3 data of GCls to search for line-like or box-shaped structures.
The GCls adopted in this work are those with the largest $J$-factors inside the HIFLUGCS catalog~\citep{Reiprich2002,Chen2007,Anderson2016}.
We use the unbinned likelihood with the sliding windows technique to search for signals and set constraints.

We search for lines and do not find any significant ($\rm TS>25$) excess, as shown in Fig.~\ref{fig:ts}.
The 42.2~GeV line candidate persists in the new P8R3 data set, even though the TS value decreases to 13.1, corresponding to a local (global) significance of $3.6\sigma$ ($2.5\sigma$).
We split the data into different event types and find the line candidate mostly comes from the EDISP2 data.
The TS value decreases to 2.3 if only the EDISP2 data are dropped.
To find which GCl contributes to the weak structure, the unbinned analysis is performed in each GCl and mild excesses are found in several GCls (Fig.~\ref{fig:ts_nsrc}).
In particular, Virgo and Ophiuchus, two galaxies with the largest boosted $J$-factors, have the highest TS values.
If the events from the first 5 GCls with the biggest boosted $J$-factors are stacked, the local (global) significance is $4.4\sigma$ ($2.7\sigma$).
However if the EDISP2 data are excluded, the local significance decreases to merely $2.9\sigma$.
We further derive the change of TS value with time (Fig.~\ref{fig:ts_lc}), which increased to the TS value of 21.2 in October 2016 but decreases since then.
There is a $\sim 2\sigma$ tension between the photon counts from the line observed in October 2016 and in March 2020.
We also searched for lines in an alternative GCl sample built from the 2MASS Redshift Survey, but no line including the tentative $\sim 42~\rm GeV$ signal is detected.
We calculate the 95\% confidence level upper limit on the DM annihilation cross section in Fig.~\ref{fig:sv}.
Only the cross section around the DM mass of 42~GeV exceeds the 95\% containment regions, however it is restricted by the observations of 2MASS GCls.
Generally speaking, since the significance of $\sim 42~\rm GeV$ line does not increase with the observing time in recent years and the parameter space is constrained by the 2MASS GCls observation, it is unlikely to have an astrosphysical origin.
The statistical fluctuation, unusual astrophysical processes (e.g.~\citet{Aharonian2012}) or incomplete understanding of energy reconstruction method of \lat might be responsible for it.

We also search for box-like structures in the baseline GCl sample.
No significant box-shaped feature is found as well (Fig.~\ref{fig:ts_box}).
A tentative excess is found at $m_\chi=84.3~\rm GeV$ (80.3~GeV, 71.2~GeV) with the maximum TS value being 12.9 (13.8, 10.6) for the boxes with degeneracy parameter $\Delta m/m_\chi=0.001$ (0.01, 0.05), which are all contributed by the sharp structure at $\sim 42~\rm GeV$ shown in the stacked SED.
But when the EDISP2 data are excluded, the TS value decreases to $\lesssim 4$.
We calculate the 95\% confidence level upper limit on the cross section of DM cascade annihilation, shown in Fig.~\ref{fig:sv_box}.
The upper limit we obtained with the averaged boost factor of $\sim 25$ is weaker than those from the Galactic center~\citep{Ibarra2012,Ibarra2013} or dSphs~\citep{LiS2018}.
However, supposing the boost factor follows the relation in~\citet{Gao2012}, GCls can be the optimal targets.

Some clues, including the low significance in EDISP1 and EDISP3 data,
the decrease of TS value with time since 2016, and the null results in the alternative GCl samples, may point towards the non-astrophysical origin of the weak line/box-like structure at $\sim 42~\rm GeV$.
However, it is still puzzling that the GCls in HIFLUGCS catalog with large $J$-factors tend to have high TS values.
A better understanding of \lat data can be helpful, but, more importantly, verifications with independent telescopes are also necessary because of the systematic uncertainties involved (see e.g.~\citet{Ackermann2013a}).
The next generation \gr missions, such as High Energy cosmic-Radiation Detection facility (HERD)~\citep{HERD2019} and Very Large Area \gr Space Telescope (VLAST),\footnote{\url{http://www2.yukawa.kyoto-u.ac.jp/~mmgw2019/slide/5th/Fan.pdf}.} will be powerful instruments to settle the problem.

\begin{acknowledgments}
    We would like to thank Kai-Kai~Duan, Shang~Li and Xiang~Li for helpful discussions.
    In the data analysis, we make use of {\tt NumPy}~\citep{numpy2020}, {\tt SciPy}~\citep{scipy2020}, {\tt Matplotlib}~\citep{matplotlib2007}, {\tt Astropy}~\citep{astropy2018} and {\tt iminuit}~\citep{iminuit} packages.
    This work is supported by the National Natural Science Foundation of China (Nos. U1738210, 12003074 and 12003069), the National Key Program for Research and Development (No. 2016YFA0400200), and the Entrepreneurship and Innovation Program of Jiangsu Province.
\end{acknowledgments}


\end{document}